\documentclass[%
 reprint,
 amsmath,amssymb,
 aps,
]{revtex4-2}

\usepackage{graphicx}
\usepackage{dcolumn}
\usepackage{bm}

\DeclareMathOperator{\Tr}{Tr}

\usepackage{xcolor}
\begin{document}

\preprint{APS/123-QED}

\title{Particle Number Fluctuations, R\'{e}nyi and Symmetry-resolved Entanglement Entropy
in Two-dimensional Fermi Gas
from Multi-dimensional Bosonization}

\author{Mao Tian Tan}
\author{Shinsei Ryu}%
\affiliation{%
 Kadanoff Center for Theoretical Physics, University of Chicago, IL 60637, USA
}%




\date{\today}

\begin{abstract}
In this paper, we revisit the computation of particle number fluctuations
and the R\'{e}nyi entanglement entropy of 
a two-dimensional Fermi gas using multi-dimensional bosonization. 
In particular, we compute these quantities for a circular Fermi surface and a circular entangling surface. Both quantities display a logarithmic violation of the area law, and the R\'{e}nyi entropy agrees with the Widom conjecture. Lastly, we compute the symmetry-resolved entanglement 
entropy for the two-dimensional circular Fermi surface and find that,
while the total entanglement entropy scales as $R\log R$, 
the symmetry-resolved entanglement 
scales as $\sqrt{R\log R}$,
where $R$ is 
the radius of
the subregion of our interest.
\end{abstract}

\maketitle


\section{Introduction}
In recent years, there has been a surge of interest 
in quantum entanglement and its various measures, 
in the condensed matter as well as the high energy physics communities
\cite{RevModPhys.82.277,LAFLORENCIE20161,Latorre_2009,RevModPhys.91.021001,Ryu_2006,Nishioka_2009,rangamani2017holographic,Headrick:2019eth,RevModPhys.90.035007}.
One of the most profound results pertaining to 
the entanglement entropy in many-body systems is the area law for ground states of gapped systems, where the entanglement entropy 
is known to be proportional to the area of a subregion \cite{RevModPhys.82.277}. 
This area law 
underlies
the simulatability of gapped ground states by matrix product states \cite{ORUS2014117}. 
Intuitively, degrees of freedom in a system with local interactions are entangled only with their neighbours,
so the entanglement entropy receives contributions primarily from the degrees of freedom situated close to the boundary.

Even though the appearance of 
the area law behaviour in ground states is ubiquitous, there are known exceptions where the area law is violated, typically with a logarithmic correction. Some well-known examples are conformal field theories in one spatial dimension, which describe quantum critical points, and 
Fermi liquids
in higher spatial dimensions with a 
Fermi surface 
\cite{Gioev_2006,PhysRevLett.96.010404}

While the von Neumann entropy and related measures are important physical quantities, they are difficult to compute analytically for generic many-body systems. 
Conformal field theories in one spatial dimension are 
among the most analytically tractable systems since the replica technique can be applied there.
In these cases, the computation of the entanglement entropy boils down to the evaluation of the correlation functions of twist operators. Another approach for one-dimensional systems would be to use the Fisher-Hartwig formula
for free systems. There are, however, fewer analytical calculations done in spatial dimensions greater than one. 
Calculations for the entanglement entropy 
of a higher dimensional Fermi surface
 can either be done by applying 
the Widom conjecture or bosonization, or by simply dividing up the multi-dimensional Fermi surface into many one-dimensional pieces where one can use the known one-dimensional results \cite{Ryu_2006, PhysRevLett.105.050502,PhysRevB.86.045109,PhysRevX.2.011012,PhysRevLett.105.050502}.

In this paper, we apply the multi-dimensional bosonization technique developed in \cite{PhysRevB.49.10877, neto1995exact} 
to calculate the entanglement entropy and related quantities of a Fermi gas analytically and non-perturbatively. Firstly, we compute the particle number cumulants generating function.
This quantity is then used to carefully derive the entanglement entropy for an isotropic Fermi gas, which is found to be in agreement with Widom's conjecture. 
This implies that the leading term in the entanglement entropy of a Fermi gas comes primarily from the modes near the Fermi surface. 
Next, the particle number cumulants generating function is also used to compute the symmetry-resolved entanglement of a two-dimensional Fermi gas \cite{PhysRevLett.120.200602}. 
We find that each particle number sector contributes an entanglement of $\sqrt{R\log{R}}$, while the total entanglement entropy scales as $R\log{R}$,
where $R$ is the radius of the subregion of our interest.

\section{Review of Multi-dimensional Bosonization}
Before proceeding with the calculations, we review 
a scheme of 
multi-dimensional bosonization 
developed in
\cite{PhysRevB.49.10877,neto1995exact}. Alternate formulations of multi-dimensional bosonization can be found in \cite{doi:10.1080/000187300243363,kopietz1997bosonization}.
Given a filled Fermi sea, we can create and annihilate particle-hole pairs with the following operators
\begin{equation}
n_{\vec{q}}(\vec{k}) = c_{\vec{k}-\frac{\vec{q}}{2}}^\dagger c^{\ }_{\vec{k}+\frac{\vec{q}}{2}},
\end{equation}
where $c_{\vec{k}}$/$c^{\dag}_{\vec{k}}$
are the electron annihilation/creation operators
with momentum $\vec{k}$.
Because they are quadratic in the fermion operators, their commutators are almost bosonic. 
However, they do not annihilate the Fermi sea. We need to normal
order the particle-hole operators relative to the Fermi sea,
so \cite{neto1995exact} defined the following operators
\begin{widetext}
\begin{align}\label{BosonOperators}
  a_{\vec{q}}(\vec{k}_F)
  &=\sum_{\vec{k}}\Phi_\Lambda(|\vec{k}-\vec{k}_F|) \left[n_{\vec{q}}(\vec{k}) \Theta(\vec{v}_{\vec{k}_F}\cdot \vec{q})+n_{-\vec{q}}(\vec{k}) \Theta(-\vec{v}_{\vec{k}_F}\cdot \vec{q})\right],
    \nonumber \\
  a_{\vec{q}}^\dagger(\vec{k}_F)
  &=\sum_{\vec{k}}\Phi_\Lambda(|\vec{k}-\vec{k}_F|) \left[n_{-\vec{q}}(\vec{k}) \Theta(\vec{v}_{\vec{k}_F}\cdot \vec{q})+n_{\vec{q}}(\vec{k}) \Theta(-\vec{v}_{\vec{k}_F}\cdot \vec{q})\right],
\end{align}
\end{widetext}
where $\Theta(x)=1(-1)$ if $x > 0 (<0)$ and $\Phi_\Lambda(|\vec{k}-\vec{k}_F|)$ is some dimensionless smearing function that keeps the vectors $\vec{k}$ close to 
the Fermi momentum $\vec{k}_F$. 
More precisely, it is defined as
\begin{equation}
\underset{\Lambda \rightarrow 0}{\lim}\Phi_\Lambda(|\vec{k}-\vec{k}_F|)= \delta_{\vec{k},\vec{k}_F}
\end{equation}
where $\Lambda$ is a momentum space cutoff. 
We have also defined the velocity of the particles as $\vec{v}_{\vec{k}}=\vec{\nabla} \epsilon_k$, with $\epsilon_{\vec{k}}$ being the spectrum of the one-particle states. The idea is to divide the Fermi
surface into patches of radius $\Lambda$ centered about $\vec{k}_F$, and
$\vec{q}$ is constrained to lie within the patch, so that $q \ll \Lambda \ll
{k}_F$.
By construction,
$a_{\vec{q}}(\vec{k}_F)$ annihilates the Fermi sea
$|F.S.\rangle$,
$
a_{\vec{q}}(\vec{k}_F)|F.S.\rangle = 0
$.

For each patch, the local density of states is 
\begin{equation}
N_\Lambda(\vec{k}_F) = \frac{1}{V} \sum_{\vec{k}} \left|\Phi_\Lambda(|\vec{k}-\vec{k}_F|) \right|^2  \delta(\mu -\epsilon_{\vec{k}}),
\end{equation}
where the chemical potential is $\mu = \epsilon_{\vec{k}_F}$ and the total system size is $V$. The total density of states is 
\begin{equation}
N(0) = \frac{1}{V}\sum_{\vec{k}}\delta(\mu -\epsilon_{\vec{k}})
\end{equation}
and they are related by $N_\Lambda(\vec{k}_F)  = \frac{N(0)}{S_d}$ for an isotropic Fermi surface,
where $S_d$ is the $d$-dimensional solid angle.
For
convenience, rescale the bosonic operators \eqref{BosonOperators}
as
\begin{equation}\label{rescaleBosons}
b_{\vec{q}}(\vec{k}_F)= \left[N_\Lambda(\vec{k}_F) V |\vec{q}\cdot\vec{v}_{\vec{k}_F}| \right]^{-1/2} a_{\vec{q}}(\vec{k}_F).
\end{equation}
These operators obey the usual bosonic algebra
\begin{equation}\label{BosonCommRel}
[b_{\vec{q}}(\vec{k}_F),b_{\vec{q}\,'}^\dagger(\vec{k}_F') ]=\delta_{\vec{k}_F,\vec{k}_F'}\left(\delta_{\vec{q},\vec{q}\,'} +\delta_{\vec{q},-\vec{q}\,'} \right).
\end{equation}

For the restricted Hilbert space that contains excitations close to the Fermi surface, the non-interacting Hamiltonian is effectively
given by
\begin{equation}\label{noninteractingHamiltonian}
H_0= \sum_{\vec{k}_F} \sum_{\vec{q}}^{\vec{q}\cdot\vec{k}_F>0} |\vec{q}\cdot\vec{v}_{\vec{k}_F}| b_{\vec{q}}^\dagger(\vec{k}_F) b^{\ }_{\vec{q}}(\vec{k}_F).
\end{equation}
We see that these bosons diagonalize the non-interacting low energy Hamiltonian. The electronic density is related to the bosons as follows
\begin{equation}\label{BosonizationIdentity}
  \rho(\vec{q})= \sum_{\vec{k}_F}^{\vec{v}_{\vec{k}_F}\cdot\vec{q}>0} \left[N_\Lambda(\vec{k}_F)  |\vec{q}\cdot\vec{v}_{\vec{k}_F}| \right]^{1/2}
  \left[b_{\vec{q}}^\dagger(-\vec{k}_F)+ b^{\ }_{\vec{q}}(\vec{k}_F) \right].
\end{equation}
This is the multi-dimensional bosonization identity which relates the fermionic density with the bosonic modes.

For the rest of the paper, we will restrict ourselves to two spatial dimensions.

\section{Free Fermion Particle Number Cumulant Generating Functional}
For a given subregion ${A}$, we define the generating function of particle number cumulants to be \cite{Calabrese_2012,PhysRevB.85.035409}
\begin{equation}\label{ParticleNumberGeneratingFunction}
\big\langle\,
e^{i \lambda \hat{N}_{A}}\,
\big\rangle,
\quad 
\lambda \in \mathbb{C},
\end{equation}
where $\hat{N}_{A}$ is the number operator 
of subregion ${A}$. The generating function produces the cumulants of the particle number distribution 
in subregion ${A}$ via
\begin{equation}\label{cumulant}
V^{(m)}_{A} = (-i \partial_\lambda)^m \log\,
\langle\, e^{i \lambda \hat{N}_{A}}\, \rangle\Big|_{\lambda = 0}.
\end{equation}
In particular, the second cumulant (the variance) is 
\begin{equation}\label{Variance}
V^{(2)}_{A} = \left\langle(\hat{N}_{A}-\langle \hat{N}_{A}\rangle)^2\right\rangle.
\end{equation}


Without interactions, the Hamiltonian is given by
(\ref{noninteractingHamiltonian}) for low lying states, so the ground state is annihilated by $b_{\vec{q}}(\vec{k}_F)$.
Defining $f(\vec{r})=i \lambda \Theta(\vec{r}\in {A})$, 
where $\Theta(\vec{r})$ 
is the two-dimensional step function,
the generating function \eqref{ParticleNumberGeneratingFunction} can be written as
\begin{align}
   \langle \,
   e^{i \lambda \hat{N}_{A}}\,
   \rangle 
   &= \Big\langle 
   \exp\Big[\int d^dr\, \rho(\vec{r})f(\vec{r}) \Big]
   \Big\rangle 
 \nonumber \\
 &= 
 \Big\langle \exp{
 \Big[\sum_{\vec{k}} \rho(\vec{k}) f(-\vec{k})\Big]}
 \Big\rangle, 
\end{align}
where the momentum-space density operator can be related to the bosonic modes via \eqref{BosonizationIdentity}. Since the expectation value is computed in the ground state of \eqref{noninteractingHamiltonian}, it can be simplified by the Baker-Campbell-Hausdorff formula. Let us further restrict ourselves to a circular Fermi surface for the rest of the paper. The generating function then simplifies to 
\begin{widetext}
\begin{equation}
   \big\langle\,
   e^{i \lambda \hat{N}_{A}}\,
   \big\rangle 
   = \exp{\left[-\frac{1}{2} \frac{\lambda^2}{V} \sum_{\vec{k}_F}
    \int\nolimits_{\vec{r}\in {A}}d^d r
    \int\nolimits_{\vec{r}\,'\in {A}}d^d r\,'
    \sum\limits_{\vec{q}}^{\vec{v}_{\vec{k}_F}\cdot \vec{q}>0} N_\Lambda(\vec{k}_F) (\vec{q}\cdot \vec{v}_{\vec{k}_F})\,
    e^{i \vec{q} \cdot(\vec{r}-\vec{r}\,')}
    e^{-\frac{\alpha |\vec{q}\cdot\vec{v}_{\vec{k}_F} |}{| \vec{v}_{\vec{k}_F} |}}\right]}. 
\end{equation}
\end{widetext}
Here, a momentum regulator $\alpha$ has been introduced so that only states with small excitation momenta normal to the Fermi surface are kept.

It remains to perform the momentum sums and the spatial integrals. 
We first integrate over the excitation momentum $\vec{q}$.
This can be done by 
decomposing $\vec{q}$
into two parts, 
$\vec{q}=\vec{q}_N+\vec{q}_T$
where 
$\vec{q}_{N/T}$
are the components normal/tangential
to the Fermi surface.
The subsequent 
spatial integral leads to
\begin{align}
    \langle\, e^{i \lambda \hat{N}_{A}}\, \rangle 
    = \exp
    \left[
    -\frac{\lambda^2}{2\pi} N N_{\Lambda}(\vec{k}_F)
    |\vec{v}_{\vec{k}_F}|
    I(R)
    \right],
\end{align}
where
$N= \sum_{\vec{k}_F}$
is the number of patches, and
\begin{align}
I(R) =
\int_{\vec{r} \in A} d^d r\, 
\frac{ \sqrt{ R^2 - y^2 } }{
R^2 - y^2 - (x+ i \alpha)^2}
\end{align}
with $\vec{r}= (x, y)$. The coordinate system was chosen such that $\vec{q}_N$ is parallel to $\hat{x}$ and $\vec{q}_T$ is parallel to $\hat{y}$.
The final result is independent of the direction of the Fermi momentum $\vec{k}_F$ due to the isotropy of the Fermi surface, so the sum over the Fermi momentum of the patches of the Fermi surface simply becomes the number of such patches. 
Evaluating the final  one-dimensional spatial integral  numerically,
we obtain the generating function for the particle number cumulants, 
\begin{equation}
\label{result 1}
    \langle\,
    e^{i \lambda \hat{N}_{A}}\,
    \rangle=\exp\left[-\frac{\lambda^2 k_F}{(2\pi)^2}\left( 2 R \ln \frac{R}{\alpha}+0.776R\right)\right].
\end{equation}
This equation is independent of the number of patches of the Fermi surface, as it should be. 
The coefficient of the logarithmic term converged  unambiguously to 2 in the numerical integration. 
With the generating function at hand, one can easily obtain the variance \eqref{Variance},
\begin{equation}
\label{SecondCumulant}
V^{(2)}_{A} = \frac{k_F}{2\pi^2} \left(2 R \ln \frac{R}{\alpha}+0.776 R\right).
\end{equation}
The leading term is simply the two-dimensional area law with a logarithmic violation, and is thus proportional to the entanglement entropy, as it should be \cite{Klich_2006,PhysRevLett.96.010404}. 
Intuitively, if the number of particles in a subregion has significant fluctuations, then there must be a large number of particles moving between the subregion and its complement, giving rise to a large amount of entanglement, so these two quantities should be proportional.

\begin{figure}
\center
\includegraphics[scale=0.4]{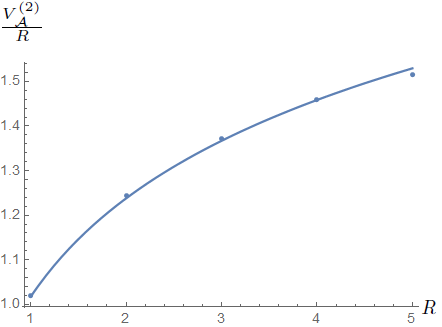}
\caption{
The variance  $V^{(2)}_{A}/R$
as a function of the radius $R$ of the subregion
obtained from the numerical evaluation of 
\eqref{VarianceOverlapMatrix} (dots)
fitted against 
the analytical result \eqref{SecondCumulant} (curve).
The fit gives $\alpha = 0.0603$. The system size is $L=20$ and $k_F = \pi$.} 
\label{NumericalVarianceFit}
\end{figure}

The variance \eqref{SecondCumulant}
can be checked with 
the following formula,
valid for free fermions systems, 
in terms of
the overlap matrix 
\cite{PhysRevLett.107.020601,Calabrese_2012,CalabreseArbitraryDim}:
%
\begin{equation}
\label{VarianceOverlapMatrix}
V_{A}^{(2)}=\Tr\, [\mathbb{A}(1-\mathbb{A})],
\end{equation}
where $\mathbb{A}$ is the overlap matrix given 
in terms of the single particle eigenfunctions
$\phi_n(\vec{r})$ as
\begin{equation}
\mathbb{A}_{nm} = \int\nolimits_{\vec{r}\in {A}} d^d r\, \phi_n^*(\vec{r})\phi_m(\vec{r}).
\end{equation}
The overlap matrix $\mathbb{A}$ is the continuum limit of the more familiar correlation function $\mathbb{C}$ used in computing entanglement entropy for free fermions on a lattice \cite{Calabrese_2012},
$
\Tr \mathbb{C}^n = \Tr \mathbb{A}^n.
$
(Working in the continuum model allows us to select a perfectly circular subregion, which is not possible in a lattice model.)
For a total system of size $L\times L$ with periodic boundary conditions,
\begin{align}
\mathbb{A}_{nm}&=\frac{1}{L^2}\int\nolimits_{\vec{r}\in A} d^2r\,  e^{i(\vec{k}_m-\vec{k}_n)\cdot\vec{r}} \\ \nonumber
&=\frac{2\pi R}{|\vec{k}_m-\vec{k}_n |L^2}J_1\left(|\vec{k}_m-\vec{k}_n |R \right),
\end{align}
where $\vec{k}$ lies within a circular Fermi surface of radius $k_F$ and $J_1(x)$ is a Bessel function of the first kind. The subregion $A$ is a disk of radius $R$ as before. 
The numerically obtained variance \eqref{VarianceOverlapMatrix} with $L=20$, $k_F = \pi$ is compared with the analytical result (\ref{SecondCumulant}) in Fig.\ \ref{NumericalVarianceFit}. 
Here, we use 
the regulator $\alpha$ as 
a fitting parameter
with
$\alpha = 0.0603$.

\section{Entanglement entropy}

In this section, we compute the R\'enyi entanglement entropy
of a free Fermi gas in a circular subregion ${A}$ with a two-dimensional isotropic Fermi surface:
\begin{equation}
S_{A}^{(n)} = \frac{1}{1-n}\log \Tr \left(\rho_{A}^n\right).
\end{equation}
As mentioned earlier,
for a free Fermi gas,
the quantum entanglement and the particle number variance for a given subregion
are proportional to each other
\cite{Klich_2006,Calabrese_2012,PhysRevB.85.035409}, 
\begin{equation}\label{CalabreseRel}
\frac{S_{A}^{(n)}}{V_{A}^{(2)}} = \frac{(1+n^{-1})\pi^2}{6}+\mathcal{O}(1),
\end{equation}
where $V_{A}^{(2)}$ is the second cumulant of the generating function we found earlier.
Using our previously obtained analytical result for the particle number variance, we readily obtain
\begin{equation}
S_{A}^{(n)}= (1+n^{-1})\frac{k_F}{6} R\ln \frac{R}{\alpha}+\cdots.
\end{equation}
The fact that we are able to obtain the leading term in the R\'{e}nyi entropy by multi-dimensional bosonization implies that the leading contribution 
comes from the modes near the Fermi surface.

Let us also mention that \eqref{CalabreseRel} is not the only way to relate the R\'{e}nyi entropy with the particle number variance. The R\'{e}nyi entropy can be written in terms of expectation values of the form \eqref{ParticleNumberGeneratingFunction} with particular choices of $\lambda$ \cite{PhysRevB.95.165101}. This approach yields the same result as \eqref{CalabreseRel}.

\section{Symmetry-Resolved Entanglement}
Having computed the R\'{e}nyi entropy, we turn our attention to the charged entanglement,
\begin{equation}\label{pathintegralwithflux}
S_{A}^{(n)}(c) =  \Tr\,
\left(\rho_{A}^n \, e^{i c \hat{N}_{A}}\right),
\quad 
c\in \mathbb{R}.
\end{equation}
This quantity has a nice interpretation of a replicated path integral with flux
insertion \cite{PhysRevLett.120.200602}, as will be demonstrated later. 
The charged entanglement entropy and 
its variants have been used to detect and distinguish symmetry-protected topological phases \cite{PhysRevB.93.195113,PhysRevB.95.045111}. It has also been studied holographically \cite{Belin:2013uta,Belin:2014mva,Pastras:2014oka,PhysRevD.93.105032}.
%
%
Performing the Fourier transform,
we obtain the symmetry-resolved entanglement \cite{PhysRevLett.120.200602},
\begin{align}
S_{A}^{(n)}(N_{A}) = \int\nolimits_{-\pi}^\pi \frac{d c}{2\pi}\, S_{A}^{(n)}(c)\, e^{-i c  N_A} 
= \Tr\, 
\left(
\rho_{A}^n \, \mathcal{P}_{N_{A}}
\right),
\end{align}
where $\mathcal{P}_{N_{A}}$ is the projector onto the subspace 
with $N_{{A}}$ particles in region $A$.
In other words, 
the symmetry-resolved entanglement is the contribution
to the $n$-th R\'{e}nyi entropy from states with $N_{A}$ particles in region $A$.

We begin by computing the charged entanglement. The following derivation generalizes the computation of 
the entanglement entropy in \cite{PhysRevB.95.165101} to compute 
the charged entanglement entropy.
Let us consider
the partial $U(1)$ rotation 
restricted to region ${A}$:
\begin{equation}
M = e^{i c \hat{N}_{A}}.
\end{equation}
In the basis of fermionic coherent states,
\begin{equation}
M = \int d\psi d\bar{\psi} d\chi d\bar{\chi}\,
e^{-(\bar{\psi}\psi+\bar{\chi}\chi)}M(\bar{\psi},\chi)|\psi\rangle\langle\bar{\chi}|,
\end{equation}
where $\psi,\bar{\psi},\chi,\bar{\chi}$ are Grassmann numbers. 
(These coherent states are constructed 
in the basis that diagonalizes the entanglement Hamiltonian.)
We have suppressed the indices of the Grassmann variables for notational simplicity. We also absorb all normalization constants into the integration measure.
Here, 
\begin{align}
  M(\bar{\psi},\chi)
&= e^{\phi \bar{\psi}\chi},
\quad 
\phi = e^{ i c}.
\end{align}

Performing the Grassmann integrals \cite{salmhofer2007renormalization}, we obtain the following simple form for the charged R\'{e}nyi entropy
\begin{equation}\label{symmresolvedpathintegral}
\Tr\, 
\left(\rho_{A}^n\,
e^{i c \hat{N}_{A}}\right)
= \int \prod_i d\alpha_i d\bar{\alpha}_i\, \rho_{A}(\bar{\alpha}_i,\alpha_i)\,
e^{\sum_{i,j}\bar{\alpha}_i T_{ij}\alpha_j},
\end{equation}
where $\alpha_i,\bar{\alpha}_i$ are Grassmann variables, and the $T$ matrix in the replica space is given by
\begin{equation}
T=
\begin{bmatrix}
      &  & & &e^{-ic} \\
    -1 &  & & &\\
      & -1 &&&\\
      &&\ddots &&\\
      &&&-1&\\
\end{bmatrix}
\end{equation}
with eigenvalues 
\begin{equation}\label{Teigenvalues}
\lambda_k = e^{i\left(\frac{2 \pi k - c}{n} \right)},
\quad
k=-\frac{n-1}{2},\cdots,\frac{n-1}{2}. 
\end{equation}
This matrix connects the fermions in each sheet of the replica path integral to the next and the phase factor in the upper right hand corner of the matrix corresponds to an Aharanov-Bohm phase that the fermions acquire if they pass through all the sheets of the replicated spacetime and go back to the original sheet. This is the reason why we can interpret the charged R\'{e}nyi entropy as a replicated path integral with flux insertion.
One can then factorize $S_{A}^{(n)}(c)$ as
\begin{equation}
S_{A}^{(n)}(c)=
\Tr\, \left(\rho_{A}^n e^{i c \hat{N}_{A}} \right) = \prod_k Z_k,
\end{equation}
where $Z_k$ is a ground state expectation value
\begin{align}
Z_k &= \langle\Psi|T_k |\Psi\rangle = \Tr \left(\rho_{A} T_k\right),
\nonumber \\
  T_k &= \exp
  \big(i\Lambda_k \sum_{j}c_j^\dagger c_j
  \big)
  = \exp\big(
  i\Lambda_k \sum_{\mu}f_\mu^\dagger f_\mu
  \big).
\end{align}
Here, $c_j$ are the real space fermions while $f_\mu$ are the fermions that
diagonalize the entanglement Hamiltonian, and they are related by a unitary
transformation.
$\Lambda_k$ is related to $\lambda_k$ as
$\Lambda_k=\frac{c-2\pi k}{n}$. 
We can now utilize our previous result for the generating function of particle number cumulants 
\eqref{result 1} with 
$\lambda=\Lambda_k$.
We thus arrive at
\begin{align}
S_{A}^{(n)}(c)&= \exp\Bigg[-\frac{k_F I(R)}{(2\pi)^2}
\sum_{k=-\frac{n-1}{2}}^{\frac{n-1}{2}}
\Lambda_{k}^2
\Bigg]
\\ \nonumber
&= 
e^{- \frac{n^2-1}{12 n} k_F I(R)}
e^{-\frac{c^2}{4 n \pi^2} k_F I(R)}.
\end{align}

The Fourier transform of $S_{A}^{(n)}(c)$ gives the R\'{e}nyi entropy for particle number $N_{A}$
\begin{align}
S_{A}^{(n)}(N_{A})=S_{A}^{(n)}(c=0)
\int\nolimits_{-\pi}^\pi \frac{d c}{2\pi}\, e^{-\frac{k_F I(R)}{n}\left( \frac{c}{2\pi}\right)^2 -i c N_{A}}.
\end{align}
Assuming $R\log R \gg 1$, 
the integrand is negligible for large $c$, 
so we might as well extend the integration region to $\mathbb{R}$ and get a Gaussian integral, leading to
\begin{equation}
S_{A}^{(n)}(N_{A})=
\sqrt{\frac{\pi n}{k_F I(R)}}e^{-\frac{k_F I(R)}{12}\frac{n^2-1}{n}-\frac{n \pi^2 N_{A}^2}{k_F I(R)}}.
\end{equation}
Finally,
the entanglement entropy for a given particle number $N_{A}$ is
\begin{align}
S_{A}(N_{A}) 
&= \frac{1}{6} \sqrt{2\pi k_F R \log{\frac{R}{\alpha}}}\, 
e^{-\frac{\pi^2  N_{A}^2}{2k_F R \log\frac{R}{\alpha}}},
\end{align}
where we kept only the leading order term in the final expression.

\section{Conclusion}

In conclusion, we applied multi-dimensional bosonization 
to compute the generating function of the particle number cumulants 
of a circular subregion $A$ for a two-dimensional Fermi gas.
This generating function is then used to compute the R\'{e}nyi entropy of the Fermi gas, which agrees with known results. These quantities show a logarithmic violation of the area law. 
We then proceed to compute the symmetry-resolved entanglement of the 2d Fermi gas, extending the results in \cite{PhysRevLett.120.200602} to two spatial dimensions. Each charge sector is observed to give a $\sqrt{R \log{R}}$ contribution to the total von Neumann entanglement entropy, which scales as $R\log{R}$. 
The success of multi-dimensional bosonization in computing these quantities suggests that one could try to apply multi-dimensional bosonization to compute other quantities in a non-perturbative ab initio approach.

While we focused on a non-interacting Fermi gas,
the power of bosonization lies in its ability to treat Fermi liquid interactions.
Before closing, we give a brief comment on the effects of Fermi liquid interactions
on the particle number cumulants. 
Utilizing the formalism of  \cite{neto1995exact},
we computed the particle number cumulants generating function
with an isotropic contact interaction
(with a spherical Fermi surface and a spherical entangling surface,
as in the case of the free fermi gas). 
In this computation, the effects of interactions 
can be incorporated by a Bogoliubov transformation, which relates the modes that diagonalize the interacting Hamiltonian to the non-interacting modes, 
thereby realizing Landau's adiabatic principle. 
Unfortunately, 
we found that the calculation is plagued by an infrared (IR) divergence.
The best way to deal with the IR divergence 
is at this moment unclear,
and left for future investigation.
We suspect that the IR divergence
might be cured with an improved treatment of collective modes within 
the bosonization framework.
Heuristically, dropping the IR divergence by hand,
we found that the coefficient of the leading logarithmic term 
decreases as we turn on interactions:
\begin{align}
&
     \big\langle\,
     e^{ i\lambda \hat{N}_A}\,
     \big\rangle
     =
     \nonumber\\
     &
     \exp
     \left[
     -\frac{\lambda^2 k_F}{2\pi^2}
     \left(
     1 - \frac{1}{2}
     \left(
     \frac{g N}{1+gN}
     \right)^2
     \right)
     R \log \frac{R}{\alpha}
     +
     \cdots
     \right],
\end{align}
where $g$ is the dimensionless coupling constant.
In particular,
it approaches half of the non-interacting value 
in the limit of infinite coupling. 
This decrease is consistent with the known result in one-dimensional 
Tomonaga-Luttinger liquids \cite{PhysRevB.85.035409}.
It would be interesting to 
calculate the particle number cumulants
numerically and compare them with the above findings.




{\it Note added}: While this draft was being prepared, \cite{Fraenkel:2019ykl} appeared
on arXiv, where the symmetry-resolved entanglement for higher-dimensional Fermi gases was computed using Widom's conjecture.

\bibliography{apssamp}

\end{document}